\documentclass[10pt,conference]{IEEEtran}
\IEEEoverridecommandlockouts
\usepackage{amsfonts}
\usepackage{cite}
\usepackage{amsmath,amssymb,amsfonts}
\usepackage{algorithmic}
\usepackage{textcomp}
\usepackage{multirow}
\usepackage{graphicx}
\usepackage[table,xcdraw]{xcolor}
\usepackage{comment}

\usepackage[frozencache]{minted}
\newtheorem{insight}{Insight}
\definecolor{LightGray}{gray}{0.97}
\def\BibTeX{{\rm B\kern-.05em{\sc i\kern-.025em b}\kern-.08em
    T\kern-.1667em\lower.7ex\hbox{E}\kern-.125emX}}

\makeatletter
\def\ps@IEEEtitlepagestyle{%
  \def\@oddfoot{\mycopyrightnotice}%
  \def\@oddhead{\hbox{}\@IEEEheaderstyle\leftmark\hfil\thepage}\relax
  \def\@evenhead{\@IEEEheaderstyle\thepage\hfil\leftmark\hbox{}}\relax
  \def\@evenfoot{}%
}
\def\mycopyrightnotice{%
  \begin{minipage}{\textwidth}
  \centering \scriptsize
  Copyright~\copyright~2023 IEEE. Personal use of this material is permitted. Permission from IEEE must be obtained for all other uses, in any current or future media, including reprinting/republishing this material for advertising or promotional purposes, creating new collective works, for resale or redistribution to servers or lists, or reuse of any copyrighted component of this work in other works.
  \end{minipage}
}
\makeatother

\begin{document}
    
\title{Towards Automated Cyber Range Design: Characterizing and Matching Demands to Supplies}

\author{\IEEEauthorblockN{Ekzhin Ear}
\IEEEauthorblockA{\textit{Department of Computer Science} \\
\textit{Uni. of Colorado Colorado Springs}\\
Colorado Springs, United States \\
eear@uccs.edu}
\and
\IEEEauthorblockN{Jose L. C. Remy}
\IEEEauthorblockA{\textit{Department of Computer Science} \\
\textit{Uni. of Colorado Colorado Springs}\\
Colorado Springs, United States \\
jcastano@uccs.edu}
\and
\IEEEauthorblockN{Shouhuai Xu}
\IEEEauthorblockA{\textit{Department of Computer Science} \\
\textit{Uni. of Colorado Colorado Springs}\\
Colorado Springs, United States \\
sxu@uccs.edu}
}

\maketitle
\begin{abstract}
Cyber ranges mimic real-world cyber environments and
are in high demand. Before building their own cyber ranges, organizations need to deeply understand what construction supplies are available to them. A fundamental supply is the cyber range architecture, which prompts an important research question: {\em Which cyber range architecture is most appropriate for an organization's requirements?} To answer this question, we propose an innovative framework to specify cyber range requirements, characterize cyber range architectures (based on our analysis of 45 cyber range architectures), 
and match cyber range architectures to cyber range requirements. 
\end{abstract}

\begin{IEEEkeywords}
cyber range, architecture, requirements, metrics
\end{IEEEkeywords}

\section{Introduction}\label{introduction_section}
{Cyber range} is an emerging technology that can aptly leverage the dynamic and informative interactions between attackers and defenders in cyberspace to mimic a real-world cyber environment. As organizations across many industry sectors have increased their technological and cyberspace footprint, 
their demand for cyber ranges have likewise increased. While there are commercial cyber ranges, there is a lack of treatment with scientific rigor, which is important because the current design of cyber ranges are heuristic or experience-based. As a consequence, customers have no choice but to {\em passively} adopt what is offered by cyber range vendors. 

In this paper, we advocate treating {cyber range} as a new kind of cybersecurity instrument and studying it with scientific rigor. This is important, for example, to help organizations choose the most appropriate cyber ranges according to their requirements, help  organizations 
customize their own cyber ranges, and provide significant insights into future cyber range design and development. 


\noindent{\bf Our Contributions}.
We make two contributions.
First, we formulate three research questions to guide the design of cyber ranges:
(i) How should we specify cyber range requirements?
(ii) How should we characterize cyber range architectures?
(iii) How should we map available cyber range architectures to given cyber range requirements?
Second, we propose an innovative conceptual framework
to systematically address these research questions,
by defining 22 attributes to specify cyber range requirements,
proposing six cyber range architectures, which we abstract from 45 real-world cyber range architectures; and
proposing an algorithm that matches cyber range architectures to satisfy an organization's needs.
    


\noindent{\bf Paper Outline}. The rest of the paper is organized as follows. Section \ref{framework_section} details the proposed framework.
Section~\ref{related_work_section} reviews related prior studies. Section~\ref{conclusion_section} concludes the paper.
\section{The Framework}\label{framework_section}
\begin{figure}[!htbp]
\vspace{-1em}
\centering{\includegraphics[width=8.5cm]{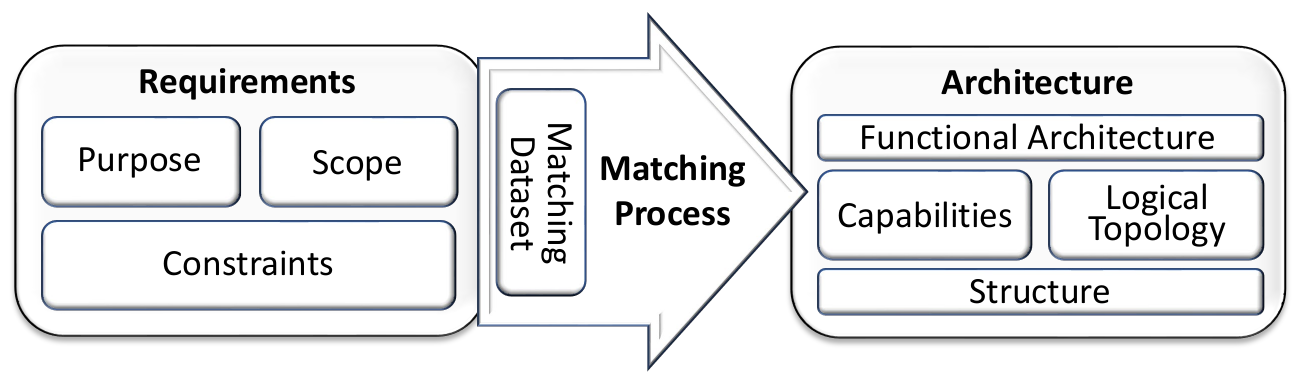}}
\vspace{-1em}
\caption{Illustration of our framework.}
\label{fig:matching_model}
\end{figure}

As highlighted in Figure \ref{fig:matching_model}, our framework has three components: (i) characterizing cyber range requirements;
(ii) characterizing cyber range architectures;
and (iii) designing an algorithm
to match cyber range architectures to cyber range requirements. These components are elaborated below.


\subsection{Characterizing Cyber Range Requirements}

\begin{figure}[!htbp]
\centering{\includegraphics[width=.45\textwidth]{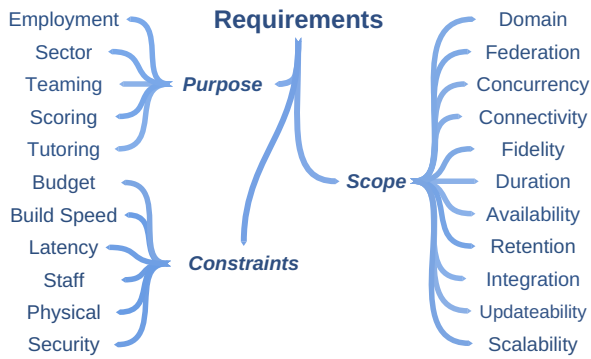}}
        \caption{Taxonomy of cyber range requirements.}
\vspace{-1em}
\label{requirements_taxonomy}
\end{figure}

Cyber range demand varies greatly in terms of organizations' requirements. 
For example, we see a preponderance of cyber ranges created for cybersecurity training purposes \cite{vykopal2021scalable, beuran2018integrated, karagiannis2022systematic, debatty2019building, potamos2021towards} and for
testing cyber-physical systems  \cite{ahmad2021design,coshatt2022design, ronkainen2012mtt}.
There are many additional use cases (such as leveraging cyber ranges as an instrument for conducting cybersecurity research) that highlight the need to accurately characterize cyber range requirements. 
Fig.~\ref{requirements_taxonomy} outlines the three requirement sets, namely {\em purpose}, {\em scope}, and {\em constraints}, with their associated attributes highlighted in Table~\ref{tbl:requirements_table} and elaborated below.

\noindent{\bf Requirement 1: Purpose}. This requirement describes the context for utilization of the cyber range and expresses its objectives in the following attributes.
\begin{itemize}
\item \textit{Employment}: This attribute captures the overall operational purpose of the cyber range. 
For example, several commercial vendors, such as Offensive Security, offer certifications through their cyber range scenarios. In this case, the purpose is to evaluate and certify candidates.
\item \textit{Sector}: This attribute identifies the sector of coverage for the cyber range, especially what types of objectives the cyber range will be used for (e.g., academic requirements can differ greatly from military ones).
\item \textit{Teaming}: This attribute identifies the participants and describes the attack-defense interactions between them. Many publicly available cyber range scenarios, such as PicoCTF and Root-Me, focus almost exclusively on the red (i.e., attacker) team, while a few others focus on blue (i.e., defender) team training. However, there is a growing demand to mimic the complex interactions of both teams.
\item \textit{Scoring}: This attribute identifies how user activity in the cyber range will be scored. This is particularly important for assessing the capabilities of red and blue team members for real-world cyber environments.
\item \textit{Tutoring}: This attribute specifies the sophistication of the instructional function of the range. This is particularly relevant to education- and training-oriented cyber ranges.
\end{itemize}

\noindent{\bf Requirement 2: Scope}. This requirement describes the extent of functionality and usage of the cyber range. Deployment size and complexity can vary significantly. For example, the Kypo Cyber Range Platform can facilitate hundreds of users, while the Kypo Cyber Sandbox Creator is for a single user via low-resourced virtual machines (VM) \cite{vykopal2021scalable}. This requirement is characterized by the following attributes.
\begin{itemize}
\item \textit{Domain}: This attribute describes the application domain to mimic, e.g., IT versus OT networks.
\item \textit{Federation}: This attribute describes the required integration with other cyber ranges, such as  integrating cyber ranges into a European cyber range ecosystem \cite{virag2021current}.
\item \textit{Concurrency}: This attribute describes the average concurrent user usage (e.g., a cyber range may be designed for use by 10 users at the same time \cite{lieskovan2021building}).
\item \textit{Connectivity}: This attribute describes the methods of user connectivity (e.g., users may require local on-site access to a cyber range vs. remote access via a gateway).
\item \textit{Fidelity}: This attribute is the degree of exactness of the cyber range to real-world systems and networks, measured by the average ratio of simulation to emulation employed. {\em Simulation} is a higher abstraction of the entity mimicked which only captures specific properties desired, while {\em emulation} mimics a more substantial set of properties, representing the entity more fully.
\item \textit{Duration}: This attribute is the continuous deployment duration required. For example, NATO's cyber range for Locked Shields requires more than 30 days in duration, even though the exercise completes  in two days \cite{smeets2022role}.
\item \textit{Availability}: This attribute describes the user time usage regime, such as whether cyber range scenarios will be deployed on-demand or will be continuously available.
\item \textit{Retention}: This attribute is the required duration of cyber range scenario data retention (e.g., a research-oriented cyber range scenario may require 6 months of retention to facilitate research investigations).
\item \textit{Integration}: This attribute identifies the internal integration between scenarios
(e.g., two red team scenarios may integrate into a new complex blue team defense scenario).
\item \textit{Updateability}: This attribute is the degree of planned updates to the infrastructure and scenario sets. For example, one cyber range design may require incremental updates to ensure its relevance to smart campus \cite{tian2018real}.
\item \textit{Scalability}: This attribute is the degree of planned increase in purpose and scope, where the degree is measured as the average increase across the range of values of \textit{purpose} and \textit{scope} attributes.
\end{itemize}

\noindent{\bf Requirement 3: Constraints}. This requirement describes constraints imposed on the cyber range designer and are associated with the attributes of 
\textit{purpose} and \textit{scope} requirements.
\begin{itemize}
\item \textit{Budget}: This attribute is the annual monetary funds available for construction, lifecycle, and maintenance.
\item \textit{Build Speed}: This attribute is the specified average maximum time allowed to provision and deploy a cyber range, and/or its scenario set.
\item \textit{Latency}: This attribute is the average permissible network and system delay time in the cyber range.
\item \textit{Staff}: This attribute is the number of full-time personnel available to support administration and scheduled/unscheduled maintenance.
\item \textit{Physical}: This attribute is the non-contiguous physical space available to house and facilitate scenarios.
\item \textit{Security}: This attribute describes the level of data and process protection required to secure the cyber range itself and to mitigate the potential damages or attacks that may originate from the cyber range against others.
\end{itemize}

\begin{table}[!htbp]
      \caption{Requirements attributes and their range of values.}
      \label{tbl:requirements_table}
      \includegraphics[width=\linewidth]{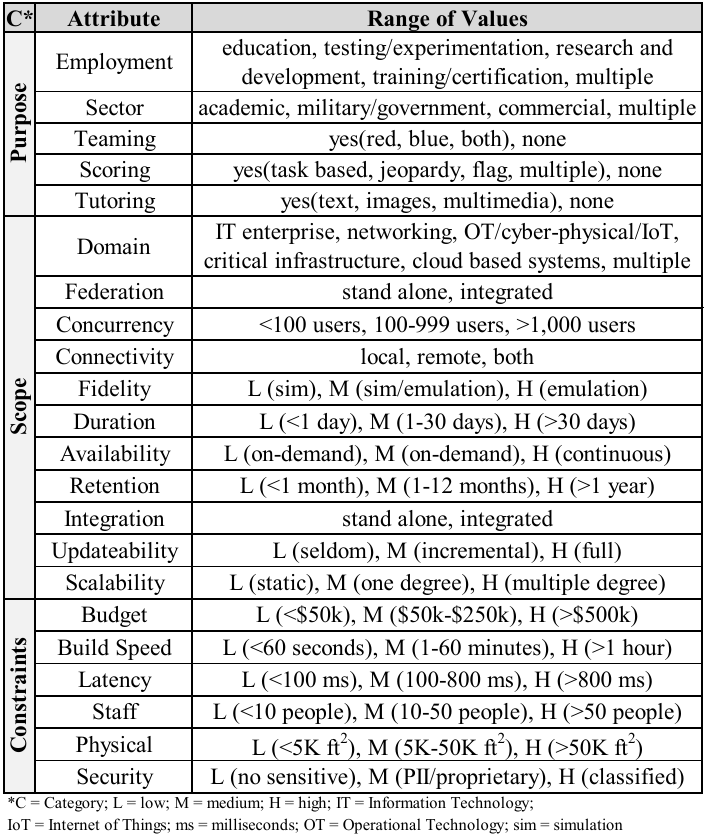}
    \end{table}

Given the preceding set of 22 attributes, where each can take multiple values, there are many combinations of cyber range characteristics. 
This means that it is not feasible to manually characterize these numerous combinations.


\subsection{Characterizing Cyber Range Architectures}

\noindent{\bf Defining Cyber Range Architectures.}
From our study of 45 architectures discussed in the literature, ranging from federated enterprise business \cite{virag2021current}, to space satellite \cite{luglio2012satellite}, to maritime systems \cite{potamos2021towards}, we define architectures (as depicted in Fig.~\ref{fig:matching_model}) to systematically describe the design of cyber ranges. 

We note that cyber ranges are typically defined by their functionality. Hence, \textit{functional architecture} sits at the top of the framework. Users interact with the cyber range at this layer. 
The \textit{functional architecture} describes the runtime environment, where assets are deployed to perform their functions (e.g., security, network, storage) and integrate with each other across a cyber range. It also describes management functions (typically found in the cantonment area of a cyber range) that allows for the deployment, maintenance, and customization of portals (i.e., user interfaces), user roles (i.e., access and permissions), environment resources (e.g., memory, storage), monitoring (e.g., network traffic), and the scenario lifecycle (i.e., creation, modification, deployment, generation, and execution). 

At the next layer, \textit{capabilities} describes the tools that enable the use and functionality of a cyber range. One difference of the cyber domain from the kinetic domain is that capabilities often come in the form of software. 
We cataloged 350 tools, including cybersecurity tools, and gained the following insight: 

\begin{insight}
\label{insight:2}
Most free/publicly available cybersecurity tools (205 of 241) are for offensive or forensics purposes.
\end{insight}

Insight \ref{insight:2} makes sense as vendors have financial incentives to produce commercial defensive risk and incident management tools. By contrast, most commercial companies are not paying for offensive cyber capabilities nor technical forensics tools. 

The \textit{logical topology} describes the various networks that support the administration and scenario deployment of the cyber range including the degree of cloud employment. In exploring potential topologies, we gained the following insight:

\begin{insight} \label{insight topology}
Within a highly managed cyber range, administrative data and processes will likely interfere with the objectives of high fidelity scenarios (e.g., experiments and certification). 
\end{insight}

Insight~\ref{insight topology} emphasizes that highly managed cyber ranges have sensors deployed across the cyber range for management, monitoring, and security, which cyber range scenario developers must account for. These hardware and software sensors, as well as the data and  traffic they collect, can introduce foreign artifacts that, for example, could corrupt the control group of an experiment, or pollute the exam network of a certification.

Lastly, the \textit{structure} defines the physical components of the cyber range. Modern hardware systems 
are complicated to configure and ensure compatibility, as well as expensive to procure, maintain, and replace. Hence, organizations must exercise significant foresight in structure design.

\noindent{\bf Designing Reference Architectures}. We propose six reference architectures. What are common to these architectures include:
(i) each architecture contains a cantonment cluster responsible for management and administration of the cyber range; 
(ii) each cantonment cluster can leverage similar tools to enable their management capabilities (e.g., syslog for monitoring); and (iii) a DMZ is used as an example security implementation, which is replaceable with other security approaches.

\begin{figure}[!htbp]
    \vspace{-1.5em}
        \centering{\includegraphics[width=.4\textwidth]{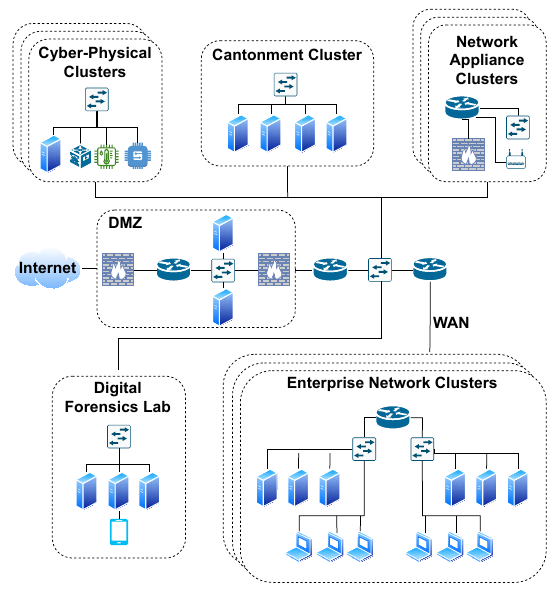}}
\vspace{-1em}
\caption{Pure physical reference architecture depicting cyber-physical sensors, network appliances for hardware testing, and hardware for digital forensics.}
\label{physical_topology}
    \end{figure}

\noindent{\it Reference Architecture 1: Pure Physical}.
The defining characteristic of this architecture is that all systems and appliances are physically the same as their real-world counterparts that they intend to mimic. Its capabilities are also the same in this regard (e.g., routing and segmentation are accomplished via physical network appliances). Fig.~\ref{physical_topology} depicts a logical topology and structure with typical use cases directly relatable to physical devices. The enterprise network clusters illustrate the potentially high resource demands of this architecture.

\begin{figure}[!htbp]
\vspace{-1em}        \centering{\includegraphics[width=7.5cm]{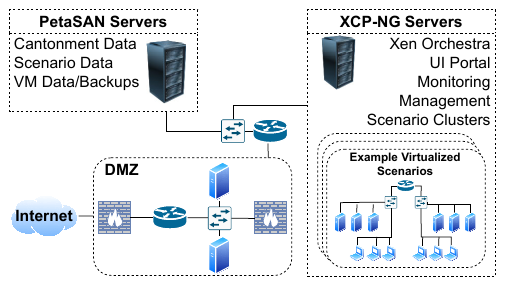}}
\vspace{-1em}
\caption{Centrally virtualized reference architecture with example capabilities (e.g., PetaSAN and XCP-ng) to enable storage pooling and virtualization.}
        \label{virtualized_topology}
    \end{figure}

\noindent{\it Reference Architecture 2: Centrally Virtualized}.
This architecture emphasizes virtualization technology over physical devices, significantly decreasing the complexity of the structure. Hence, tools like Xen Server, KVM, and VirtualBox are key capabilities to provide very high fidelity emulation.  Fig.~\ref{virtualized_topology} depicts the logical/physical topology where entire enterprise network clusters are virtualized within a server stack, as well as all cantonment cluster functionalities.

\noindent{\it Reference Architecture 3: On-Premise Cloud}.
Cloud properties (e.g., self service capable, resource pooling, rapidly elastic) make this architecture distinct, while its logical/physical topology is quite similar to Fig.~\ref{virtualized_topology} because private clouds are enabled by virtualization (e.g., via XCP-ng). Openstack is the main capability to create a private cloud. However, this comes with immense complexity, comprising of over 30 services, countless configurations files, and at least a dozen nodes. 

\noindent{\it Reference Architecture 4: Public Cloud}.
The public cloud approach benefits from cloud properties while offloading cloud maintenance tasks to third parties that have seemingly limitless resources (e.g., compute), though at a price. Further, the management plane provides robust and mature capabilities. It also natively supports automation tools like Terraform, though core cloud capabilities are proprietary. Its topology is also similar to Fig.~\ref{virtualized_topology}, except it resides outside the organization.

\begin{figure}[!htbp]
\vspace{-1em}    \centering{\includegraphics[width=.48\textwidth]{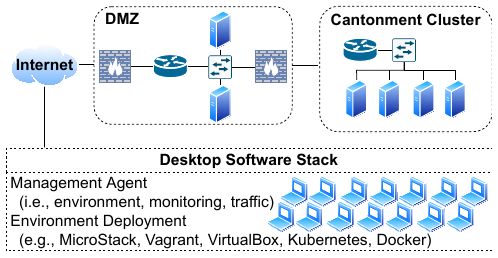}}
\vspace{-1em}
\caption{The distributed virtualization reference architecture where low resource remote nodes can host VMs, containers, and/or micro-clouds.}
        \label{distributed_topology}
    \end{figure}

\noindent{\it Reference Architecture 5: Distributed Virtualization}.
This approach emphasizes the use of remote physical nodes (e.g., end-user desktops) to house cyber range scenarios. While a cantonment cluster centrally manages the overall cyber range, more control resides with the end-user (i.e., root user of the desktop). Fig.~\ref{distributed_topology} illustrates the logical/physical topology. A use case is a classroom of students who only interact with a few VMs in simple scenarios. Virtualization, micro-cloud, containerization, and automation tools such as VirtualBox, MicroStack, Docker, and Vagrant are key capabilities.

\begin{figure}[!htbp]
    \vspace{-1em}
        \centering{\includegraphics[width=.49\textwidth]{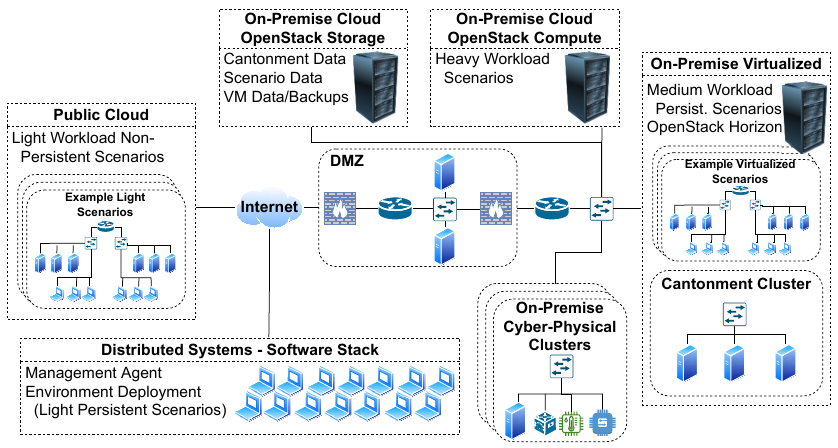}}
        \vspace{-1em}
        \caption{Hybrid reference architecture incorporating other architectures.}
        \label{hybrid_topology}
    \end{figure}

\noindent{\it Reference Architecture 6: Hybrid}.
This architecture leverages the strengths and capabilities of the other approaches, bringing them together into a single architecture of varying densities. Fig.~\ref{hybrid_topology} depicts the architecture where: a public cloud hosts low-resourced, non-persistent VMs (decreasing subscription cost); a private cloud hosts persistent, data-rich, and heavy workload scenarios (leveraging cloud properties, e.g., resource pooling); an on-premise centrally virtualized environment handles medium persistent workloads (decreasing staffing cost); distributed virtualization facilitates light persistent scenarios (decreasing cost and administration); and an on-premise physical network supports cyber-physical testing (supporting required objectives). However, this approach has the drawback of increasing the overall complexity of the architecture.

Through our design of the six reference architectures, we gained the following insight:


\begin{insight}\label{insight patching}
    It is best to employ well-maintained infrastructure capabilities and not inject custom code patching.
\end{insight}

Insight~\ref{insight patching} speaks to the complexity of cyber range infrastructures. Well-supported hardware and software capabilities (i.e., by regularly issued vendor patches) allow the cyber range to grow and mature with new innovations. Conversely, customizing cyber range deployments by directly patching code (e.g., in the XCP-ng codebase) has a greater risk of breaking functionality and snowballing overhead requirements. 

\noindent{\bf Defining Metrics to Characterize Cyber Range Architectures}.
Appropriate characterization of cyber range architectures prior to technical design and procurement saves money and time. This prompts us to define four categories of metrics in light of the requirement attributes we have discussed.

\noindent{\it Metric Category 1: Scope}.
In consideration of the \textit{scope} of the cyber range demand, we define two metrics.
\textit{Extensibility} is the degree of ability to shift across the possible values of the various \textit{scope} requirement attributes (e.g., from a pure IT enterprise domain to cyber-physical).
\textit{Capacity} is the degree of ability to achieve the highest value levels across the various \textit{scope} requirement attributes (e.g., $>1,000$ concurrent users).

\noindent{\it Metric Category 2: Performance}.
Concerning the \textit{performance} of the cyber range, we define two metrics.
\textit{Build Speed} is the amount of time (in seconds, minutes, or hours) it takes to provision and deploy a scenario set.
\textit{Latency} is the amount of average network and system delay time (in milliseconds).

\noindent{\it Metric Category 3: Cost}.   
Likewise, the \textit{cost} of the cyber range is defined in two metrics.
\textit{Budget} is the amount of annual monetary funds required to construct and maintain the cyber range.
\textit{Staff} is the number of personnel required to support administration and maintenance of the cyber range.

\noindent{\it Metric Category 4: Security}. 
We propose using standard 
\textit{security} metrics: confidentiality, integrity, availability, non-repudiation, authenticity, and privacy. 


\noindent{\bf Analyzing Reference Architectures}.
We apply these metrics at the ordinal scale and qualitatively via a Likert Scale approach \cite{batterton2017likert}.
We compare the reference architectures (of a sizeable deployment) to an average cyber range deployment scenario (i.e., to a medium-sized organization's enterprise business environment).
Table~\ref{tbl:analysis} summarizes our analysis, enumerating our qualitative results and the significant strengths and weaknesses of each architecture per metric category.



    \begin{table*}[!htbp]
      \caption{Qualitative analysis of our reference architectures, annotated with applicable strengths (+) and weaknesses (-).}
      \label{tbl:analysis}
      \includegraphics[width=\textwidth]{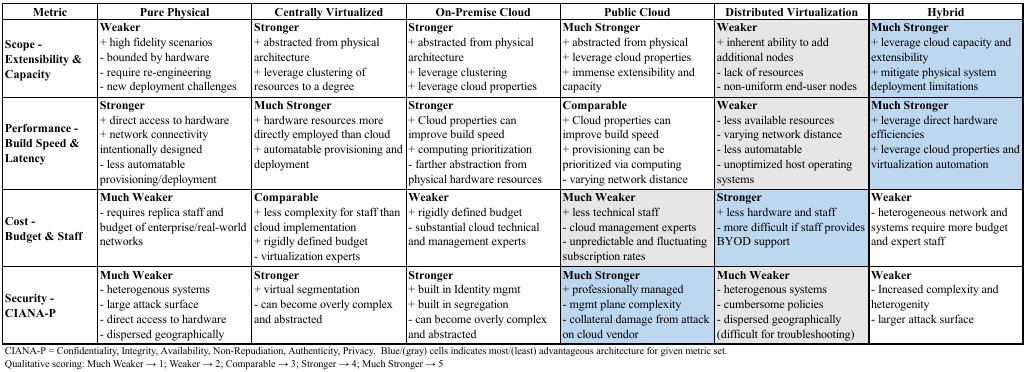}
      \vspace{-3em}
    \end{table*}

\subsection{Matching Cyber Range Architectures to Requirements}

\noindent{\bf Building a Matching Dataset}.
This dataset contains the matching of each requirement attribute to each of the six architectures.
We make the following definitions:
(i) {\it Supportability}: the level of ability a reference architecture is able to facilitate a requirement attribute. 
(ii) {\it Significance}: the level of importance of a requirement attribute given its value (e.g., significance of a {\it budget} value of {\it low} is different than {\it medium}).
We apply our domain expertise 
to assess supportability by qualitatively scoring each architecture per attribute, and significance by providing weights for each attribute per value (where the range of values are enumerated in Table~\ref{tbl:requirements_table}). We then store the resulting dataset as a CSV with the following columns: 
(i) {\it attribute name};
(ii) {\it attribute value};
(iii) {\it attribute weight}: the weight assigned to the attribute value;
(iv) {\it architecture scores}: for the remaining six columns, the assigned supportability score for each reference architecture.



\noindent{\bf Automating the Matching Process}.
We automate the matching process (as depicted in Fig~\ref{fig:matching_model}) via a python script that ingests the {\it matching dataset} and requirements of interest. It outputs a score for each reference architecture, along with a heat map to visualize and explain the score. At a high level, the score reflects to what extent an architecture satisfies the given cyber range requirements; the architecture with the highest score can be used to build a cyber range. Core functions of the script are given below:




\begin{minted}
[
frame=lines,
%framesep=2mm,
breaklines,
xleftmargin=\parindent,
baselinestretch=1.2,
%bgcolor=LightGray,
fontsize=\scriptsize,
linenos
]
{python}
def score_lookup(user_input, matching_dataset):
    for key, value in user_input:
        select line from matching_dataset where (attribute == key and value == value)
        append line to score_lookup_df
return score_lookup_df
def score_calculation(score_lookup_df):
    for architecture column in score_lookup_df:
        append sum(score_looup_df.weight * architecture.score) to architecture_scores_df
return architecture_scores_df
\end{minted}
The {\it score\_lookup} function retrieves the applicable CSV line from the {\it matching dataset} for every requirement attribute value selected by the user and provides it to the {\it score\_calculation} function, which
computes the score for each reference architecture by summing the products of each requirement attribute value's weight and architecture score. It outputs the score totals and provides this computed data to a heat map generation function.

\begin{figure}[htbp]
\vspace{-1em}        \centering{\includegraphics[width=.48\textwidth]{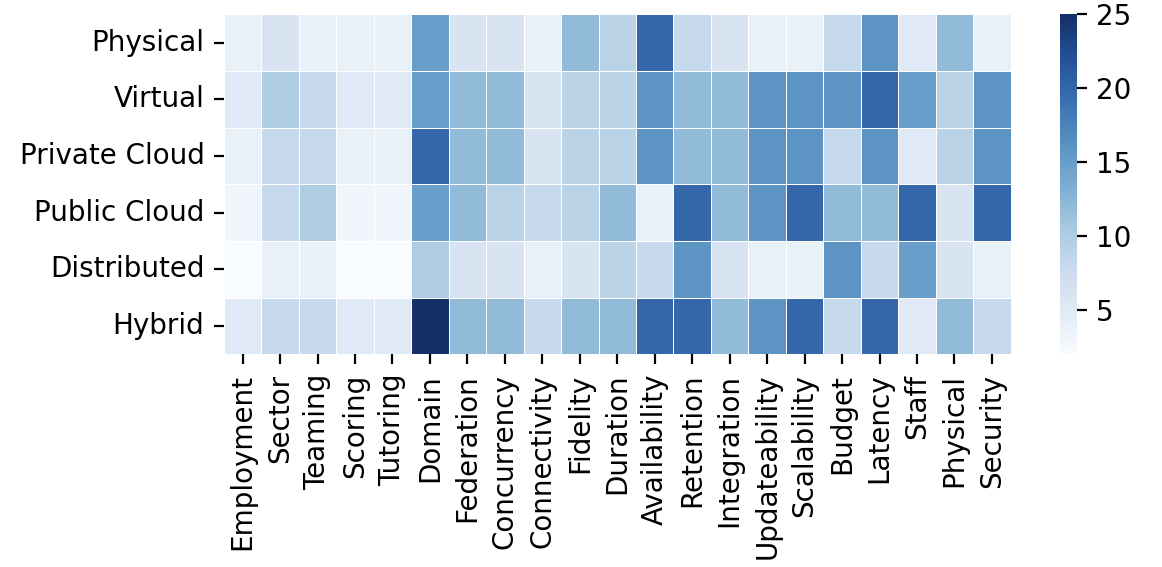}}
\vspace{-1em}
\caption{Heat map depicting a sample result, where the hybrid architecture is recommended especially because of the strong match in terms of domain, availability, retention, scalability, and latency requirements.}
        \label{matching tool screenshot}
\end{figure}

Fig.~\ref{matching tool screenshot} shows a sample output with reference architectures indicated in the $y$-axis and requirement attributes indicated in the $x$-axis. At a high level, the heat map provides explainability to the result by visualizing how requirements have impacted the selection of reference architectures. 





\section{Related Work}\label{related_work_section}

We divide related prior studies into three categories.
{First, to catalog cyber ranges, there are studies on 
characterizing cyber range functions \cite{yamin2020cyber, ukwandu2020review}, on conducting questionnaire-based survey of cyber range components and tools \cite{chouliaras2021cyber}, on recommending  
cyber ranges according to 
objectives \cite{davis2013survey}, and on summarizing cyber range infrastructures and capture-the-flag environments \cite{kucek2020empirical}. The present study 
goes much beyond these studies in terms of comprehensiveness, especially 
concerning architectures.}
Second, there are studies on creating and managing 
cyber ranges \cite{orbinato2021next}, on modeling and detecting flaws in cyber range-based training \cite{macak2022applying}, on qualitatively analyzing a particular cyber range \cite{priyadarshini2018features}, and on assessment methodologies for cybersecurity training platforms \cite{beuran2023capability}. 
The present study goes beyond them by proposing a systematic set of attributes and metrics (e.g., extensibility and capacity)
for assessing and characterizing cyber range architectures.
Third, there are studies on extending on-premise cyber ranges to public \cite{beuran2022aws} and private cloud infrastructures \cite{luise2021demand}, and on combining virtualized and physical devices into a hybrid cyber range \cite{farhat2021design}.
The present study 
goes beyond these studies by proposing a systematic set of cyber range reference architectures.

\section{Conclusion for Future Work}\label{conclusion_section}
We have presented a framework for specifying cyber range requirements, characterizing cyber range architectures, and matching cyber range architectures to requirements. This study represents a significant first step towards automating the design of cyber ranges according to requirements.

The present study has several limitations which need to be addressed in the future. 
First, the matching algorithm heavily relies on our domain expertise; a more intelligent algorithm is needed.
Second, we used qualitative metrics to assess the reference architectures; future study needs to incorporate quantitative metrics (e.g., \cite{Pendleton16,XuSTRAM2018ACMCSUR,XuSciSec2021SARR,XuACMCSUR2021ArmsRace,XuMTD2020,XuTDSC2021DynamicDivresity}).
Third, we focused on assessing general reference architectures; we want to explore more industry- and sector-specific architectures in the future.

\noindent\textbf{Acknowledgement}. This work was supported by NSF Grants \#2122631 and \#2115134, and Colorado State Bill 18-086. 

\bibliographystyle{ieeetr}
\bibliography{references,metrics}

\end{document}